\documentclass{article}

\usepackage[final]{nips_2018}

\usepackage[utf8]{inputenc} % allow utf-8 input
\usepackage[T1]{fontenc}    % use 8-bit T1 fonts
\usepackage{hyperref}       % hyperlinks
\usepackage{url}            % simple URL typesetting
\usepackage{booktabs}       % professional-quality tables
\usepackage{amsfonts}       % blackboard math symbols
\usepackage{nicefrac}       % compact symbols for 1/2, etc.
\usepackage{microtype}      % microtypography

\usepackage{comment}
\usepackage{color}
\usepackage{cite}
\usepackage{amsmath, amstext, amssymb,bm}
\usepackage{array}
\usepackage{stfloats}
\usepackage{url}
\usepackage{graphicx}
\usepackage{algorithm}
\usepackage[noend]{algpseudocode}
\usepackage{subcaption}

\usepackage{tikz}
\usetikzlibrary{%
	arrows,
	positioning
}

\title{From Deep to Physics-Informed Learning of Turbulence: Diagnostics}

\author{
	Ryan King \thanks{National Renewable Energy Laboratory, Golden, CO, USA (e-mail: Ryan.King@nrel.gov)}
	\And
	Oliver Hennigh \thanks{Los Alamos National Laboratory, X Computational Physics-4 \& CNLS,  Los Alamos, NM 87545, USA (e-mail: hennigo@lanl.gov)}
	\And
	A. Mohan \& M. Chertkov \thanks{Los Alamos National Laboratory, Theoretical Division T-4 \& CNLS, Los Alamos, NM 87545, USA (e-mails: \{arvindm,chertkov\}@lanl.gov)}
}
	
\begin{document}
	
\maketitle

\begin{abstract}
We describe tests validating progress made toward acceleration and automation of hydrodynamic codes in the regime of developed turbulence by three {\bf Deep Learning} (DL) Neural Network (NN) schemes trained on {\bf Direct Numerical Simulations} of turbulence. Even the bare DL solutions, which do not take into account any physics of turbulence explicitly, are impressively good overall when it comes to qualitative description of important features of turbulence. However, the early tests have also uncovered some caveats of the DL approaches. We observe that the static DL scheme, implementing Convolutional GAN and trained on spatial snapshots of turbulence, fails to reproduce intermittency of turbulent fluctuations at small scales and details of the turbulence geometry at large scales.  We show that the dynamic NN schemes, namely LAT-NET and Compressed Convolutional LSTM, trained on a temporal sequence of turbulence snapshots are capable to correct for the caveats of the static NN.  We suggest a path forward towards improving reproducibility of the large-scale geometry of turbulence with NN. 
\end{abstract}

%\vspace{-1cm}
\section{Introduction}
\label{sec:intro}
%\vspace{-0.5cm}

In this manuscript, reporting first results  of the approach focusing on developing  a Physics-Informed Machine Learning (PIML) framework to improve turbulence model implementations in hydrodynamic codes, see also the Supplementary Material (SM) A, we aim to verify whether various statistical properties, constraints, and relations not enforced explicitly within the DL training on the ground truth, Direct Numerical Simulation (DNS), data hold. To do this, we compare results extracted from the training data and from the generated/synthetic data. Three DL schemes, GAN of \citep{King2017},  LAT-NET of \citep{Hennigh2017}, generalized to arbitrary solvers, and newly developed Compressed Convolutional Long Short-Term Memory (CC-LSTM) scheme are juxtaposed within the setting of the homogeneous, isotropic, stationary turbulence. Specifically, we have verified (a) spectrum of energy fluctuations over scales, (b) statistics of velocity gradient (small/viscous scale object), (c) anomalous exponents of higher order velocity increments, and (d) 
statistics of the coarse-grained velocity-strain alignment represented in the plane of the coarse-grained velocity gradient tensor. Logic behind and details of the diagnostics, suggested in our prior work on the subject \citep{King2017}, are described in the SM B. Static GAN DL scheme is trained on the spatial data (instantaneous snapshots) from the Johns Hopkins turbulence database \citep{JH_data}. The dynamic schemes, LAT-NET and CC-LSTM, are trained on the dynamic (sequence of snapshots) data from the SpectralDNS code \citep{spectralDNS}. Description of the DL schemes and results of applying the two step diagnostics to the the DL schemes are described in Section \ref{sec:DLT}. We draw conclusions and suggest a path forward in Section \ref{sec:conclusions}.

\section{Deep Learning for Turbulence}
\label{sec:DLT}

\begin{figure}
\centering
		\begin{subfigure}{0.23\textwidth}
		\includegraphics[width=\textwidth]{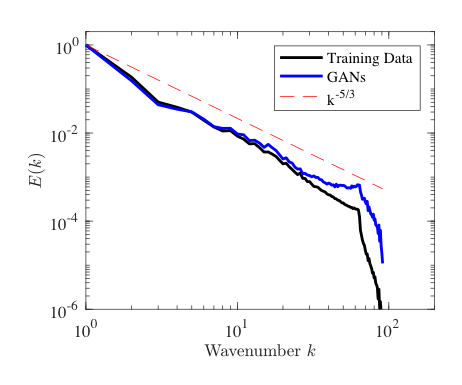}
		\caption{Energy spectra - testing all (inertial range) scales.}
		\end{subfigure}
		\begin{subfigure}{0.23\textwidth}
		\includegraphics[width=\textwidth]{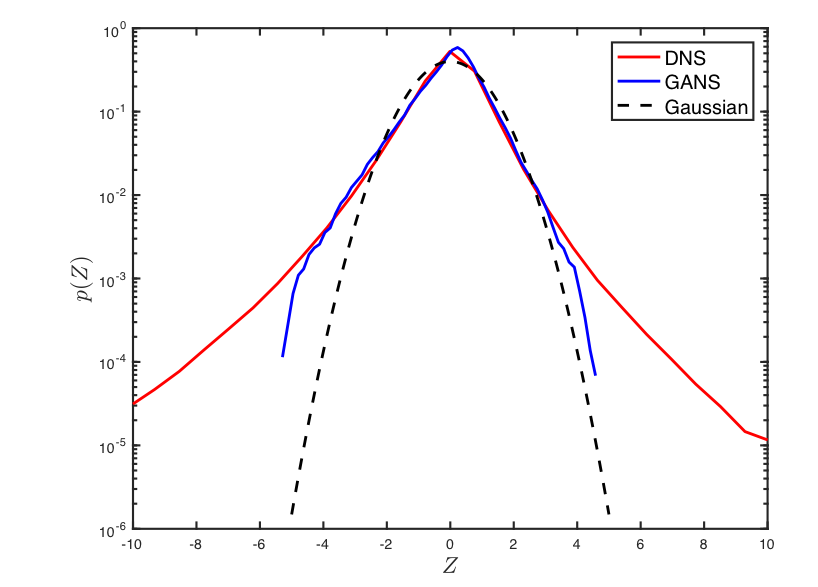}
		\caption{Intermittency of velocity gradients -- testing the smallest scales.}
		\end{subfigure}
%		\\
		\begin{subfigure}{0.23\textwidth}
		\includegraphics[width=\textwidth]{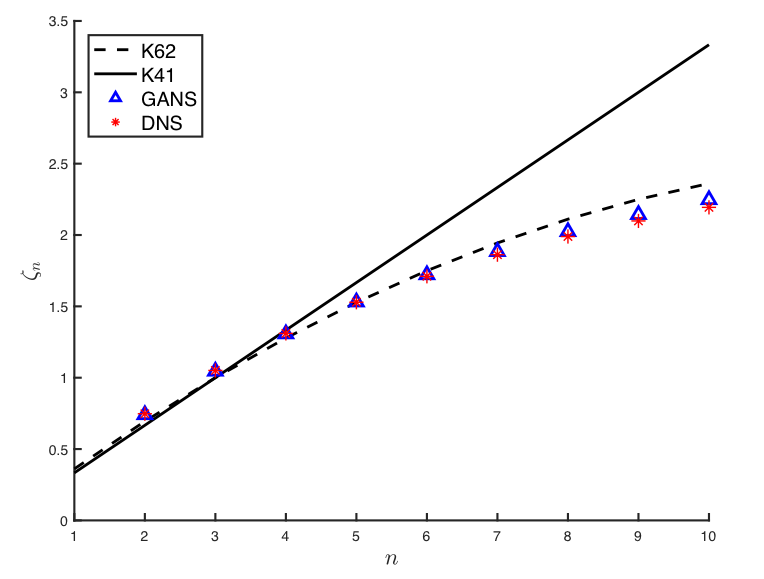}
		\caption{Anomalous scaling of structure functions -- testing intermittency at the intermediate (inertial range) scales.}
		\end{subfigure}
		\begin{subfigure}{0.28\textwidth}
		\includegraphics[width=\textwidth]{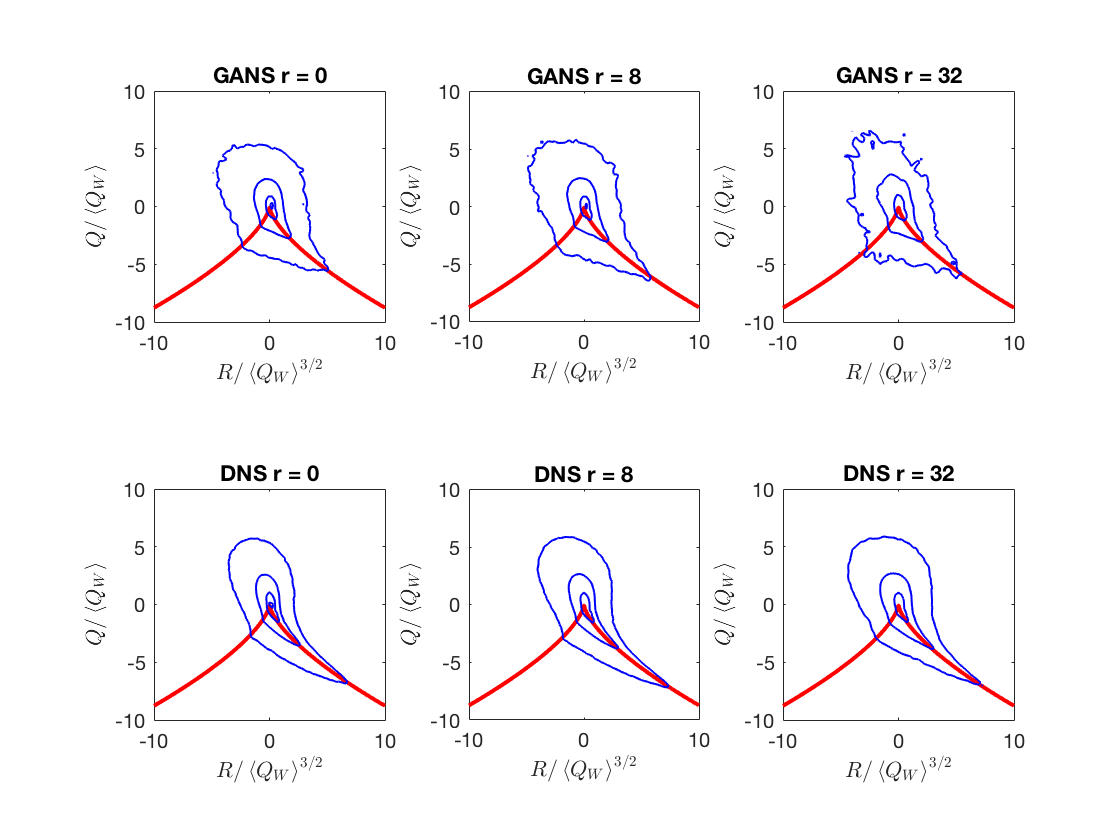}
		\caption{Coarse-grained Q and R joint PDF -- testing details details of the turbulence flow geometry (vorticity-strain alignment) at different scales.}
		\end{subfigure}
		\caption{Turbulence diagnostics showing the GANs output captures many characteristics of real turbulence.  In panel (a) the GANs output preserves the kinetic energy spectra of the training data except for a bit of extra energy at the smallest length scales.  Panel (b) shows a PDF of the normalized velocity gradients that correctly captures the non-Gaussian and negative skewness that is known to characterize intermittency in turbulence.  Panel (c) shows the GANs correctly captures anomalous scaling exponents of higher order structure functions.  Finally, panel (d) shows the classic teardrop shape of the joint PDF of the Q and R invariants of the velocity gradient tensor, testing details of flow geometry \citep{99CPS}.}
		\label{fig:GANS}
\end{figure}

\emph{Generating Turbulence through Adversarial Training.} Recent advances in DL have proven remarkably successful on image-based problems such as generation, classification, and denoising.  Much of the success of these DL techniques relies on the hierarchical identification and abstraction of features present in images using deep networks.  A complex superposition of structures also occurs in turbulent flows, naturally leading us to examine whether techniques developed for images can be used to learn turbulent flow physics. It was demonstrated in \citep{King2017} that Generative Adversarial Networks (GANs) can be trained on high fidelity DNS to generate high-quality synthetic images of turbulent flows.  GAN was originally introduced by Goodfellow et al, \citep{Goodfellow2014}, and relies on competition between two deep Neural Networks (NNs).  The first network, the generator, attempts to create synthetic images that appear to be drawn from the correct high dimensional distribution.  The second network, the discriminator, compares the proposed images against a set of training data and estimates the likelihood of the image being genuine.  A min-max optimization problem is solved iteratively to train both networks' parameters as they compete against each other.  Over time, the generator learns to draw more plausible images while the discriminator's ability to distinguish real and synthetic images also improves. Our GAN implementation makes use of the Convolution (C-GAN) for both the generator and discriminator using the architecture guidelines for stable deep C-GANs recommended by Radford et al \citep{Radford2016} and Reed et al \citep{2016Reed}.  We train on 2D slices of 3D homogeneous isotropic turbulence from the Johns Hopkins Turbulence Database \citep{JH_data}.  The three velocity components map naturally to different red, green, and blue color channels in imaged-based deep learning networks.  We calculated a number of common turbulence quantities on the GANs output as part of our survey of DL techniques.  The results, summarized in Figure \ref{fig:GANS}, suggest that the generator has successfully learned to sample from the full multi-point Probability Distribution Functions (PDFs) of valid turbulent flow fields.

\begin{figure}[!h]
\centering
    \begin{subfigure}{0.23\textwidth}
       \hspace{-2.5cm} 
       \includegraphics[width=1.75\textwidth,page=2]{Figures/LatNet_fig_v2.pdf}
	   %\caption{LAT-NET training and evaluation.} 
	\end{subfigure}
	\begin{subfigure}{0.23\textwidth}
	\hspace{-1cm} 
		\includegraphics[width=0.9\textwidth]{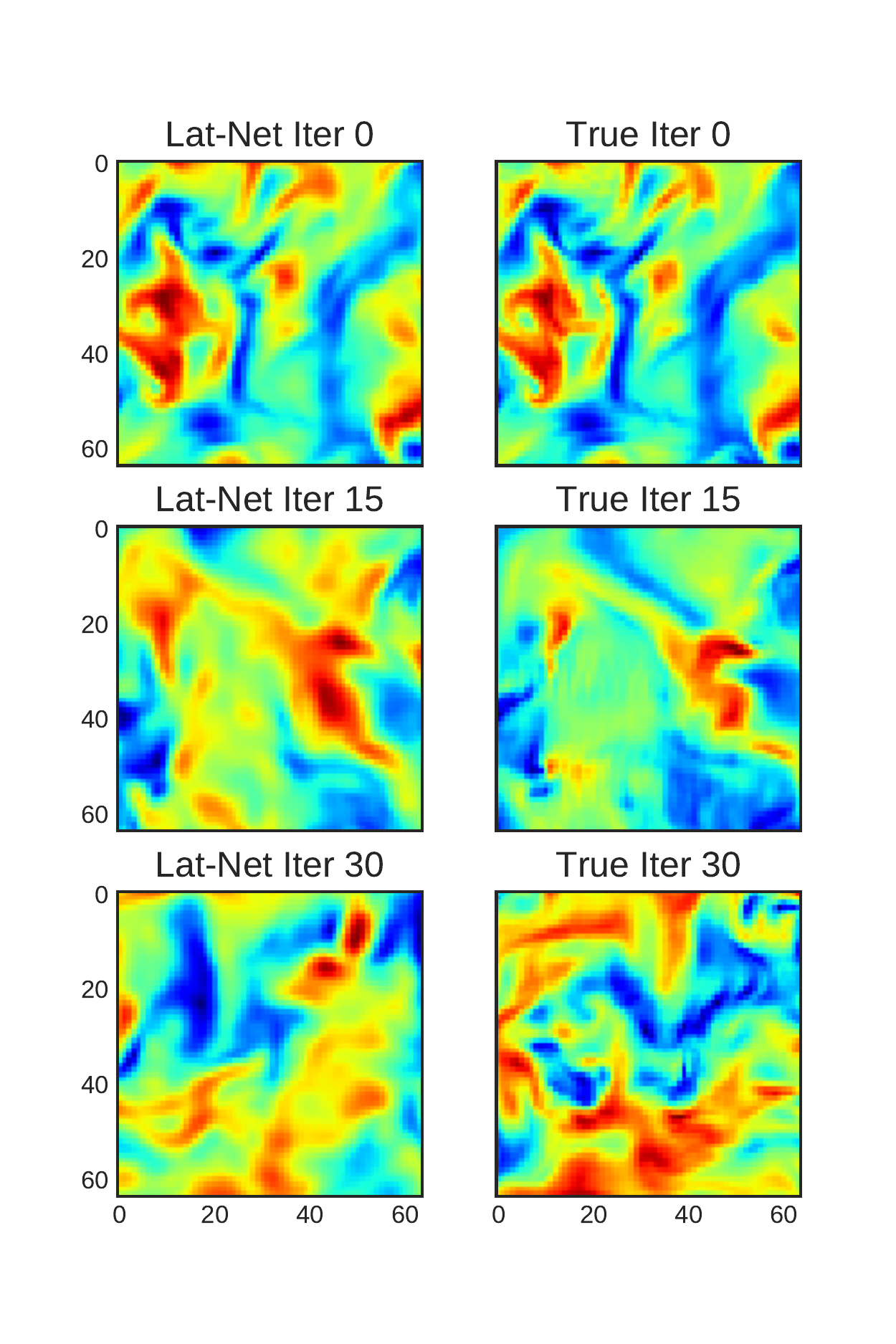}
	\end{subfigure}
	\begin{subfigure}{0.23\textwidth}
       \hspace{-1cm} 
       \includegraphics[width=1.25\textwidth,page=2]{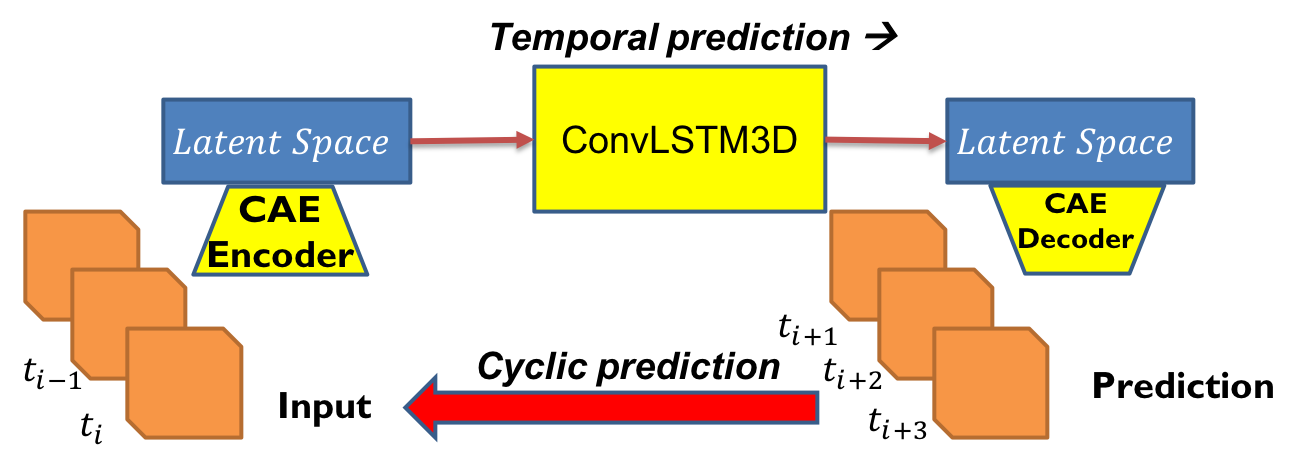}
	\end{subfigure}
	\begin{subfigure}{0.23\textwidth}
		\includegraphics[width=\textwidth]{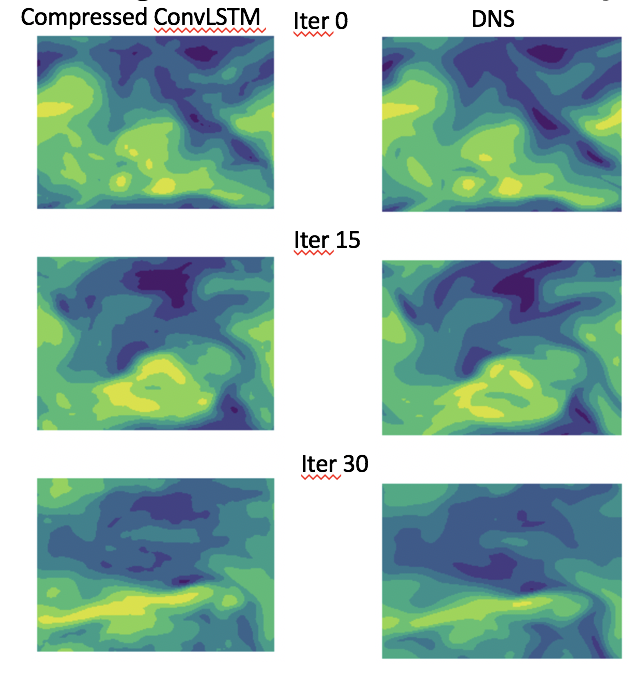}
	\end{subfigure}
	\caption{LAT-NET (left) \& CC-LSTM (right): (a) training and evaluation \& 
	(b) visualization of predicted flow.}
	\label{fig:LatNet+CC-LSTM}
\end{figure}

\begin{figure}[!h]
\centering
		\begin{subfigure}{0.18\textwidth}
		\includegraphics[width=\textwidth]{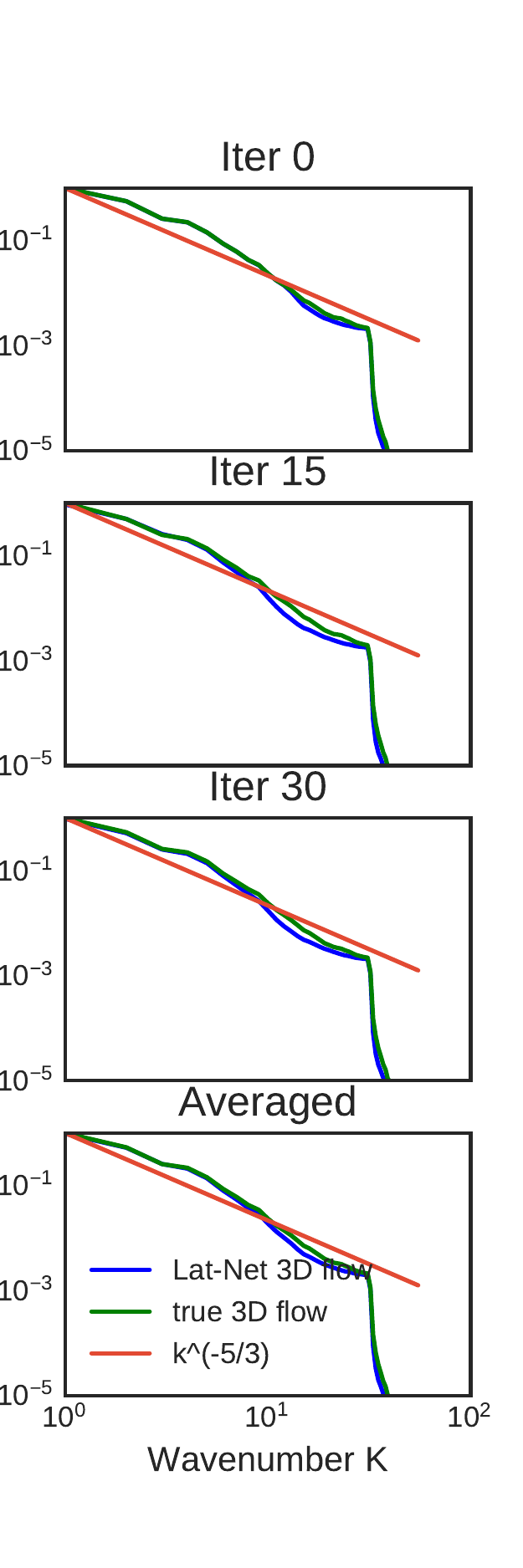}
		\caption{Energy spectra.}
		\end{subfigure}
		\begin{subfigure}{0.18\textwidth}
		\includegraphics[width=\textwidth]{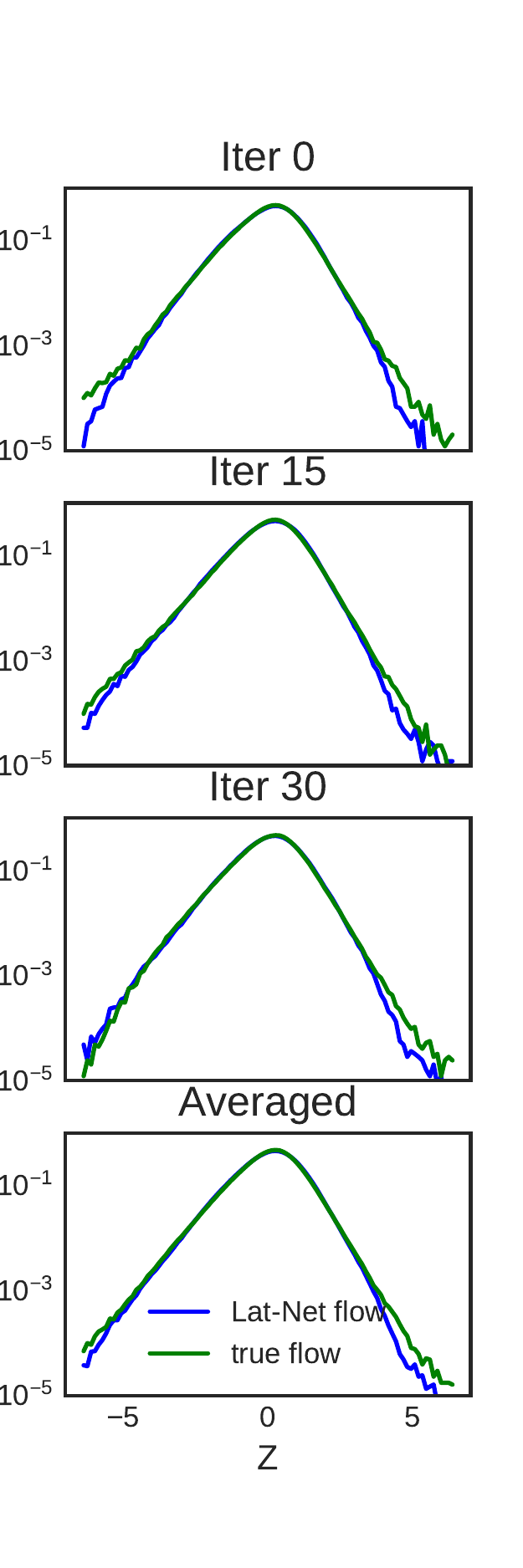}
		\caption{Intermittency.}
		\end{subfigure}
		\begin{subfigure}{0.18\textwidth}
		\includegraphics[width=\textwidth]{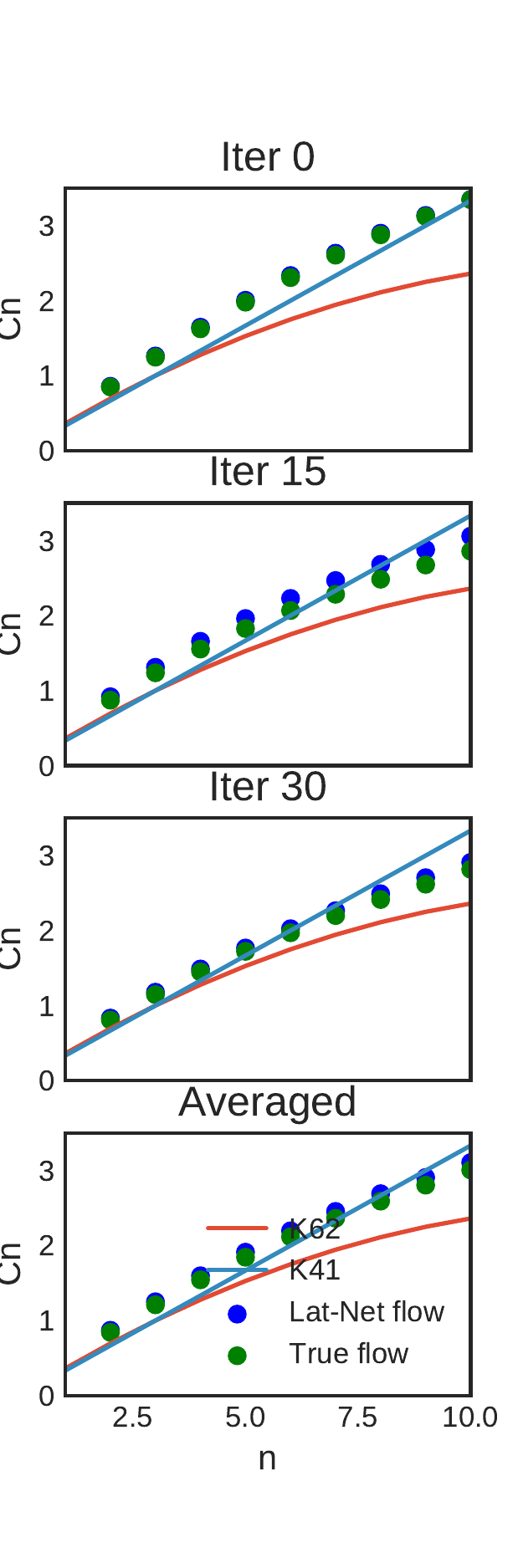}
		\caption{Anomalous Scaling.}
		\end{subfigure}
		\begin{subfigure}{0.36\textwidth}
		\includegraphics[width=\textwidth]{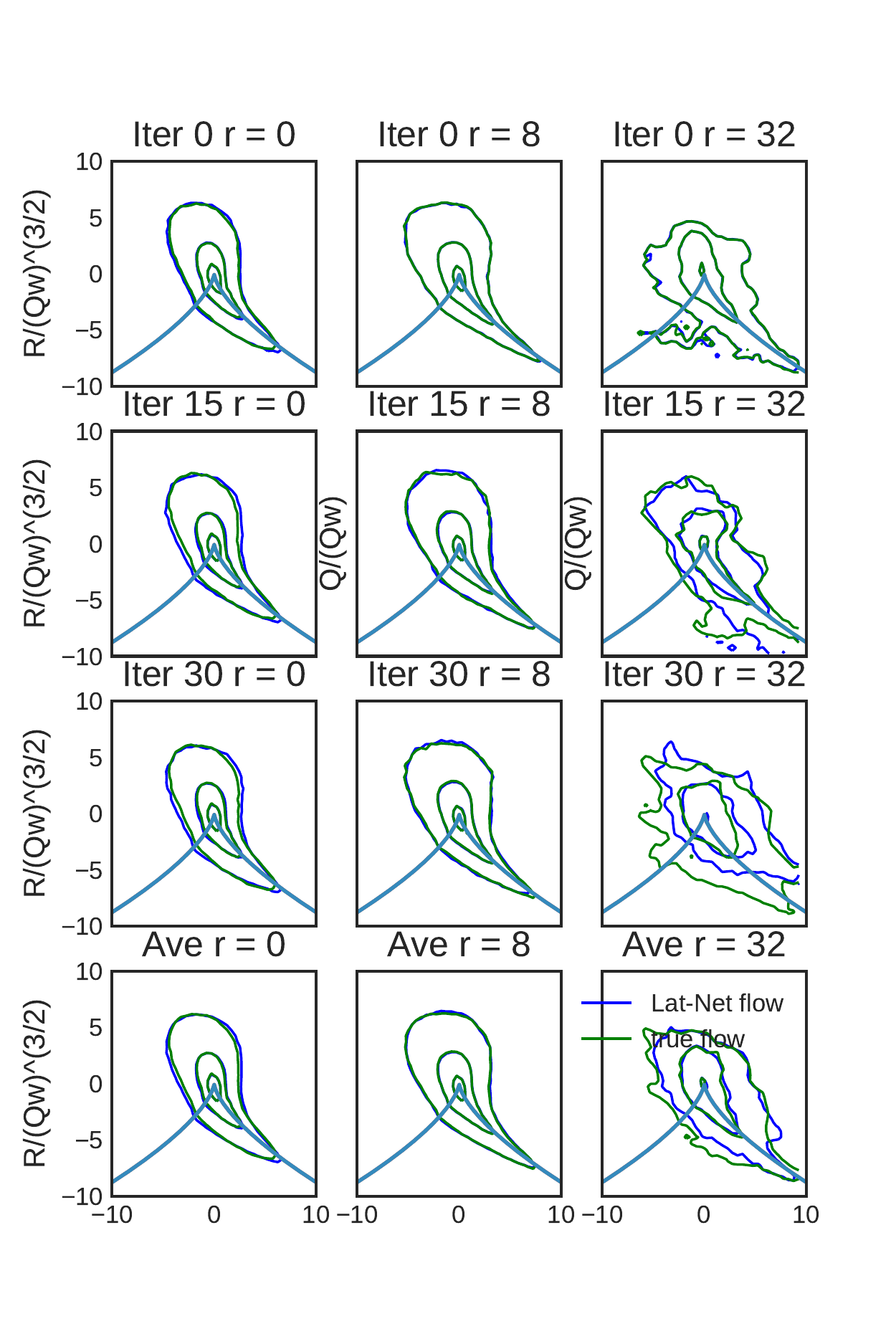}
		\caption{Coarsed-grained Q and R joint PDF's.}
		\end{subfigure}
		\begin{subfigure}{0.18\textwidth}
		\includegraphics[width=\textwidth]{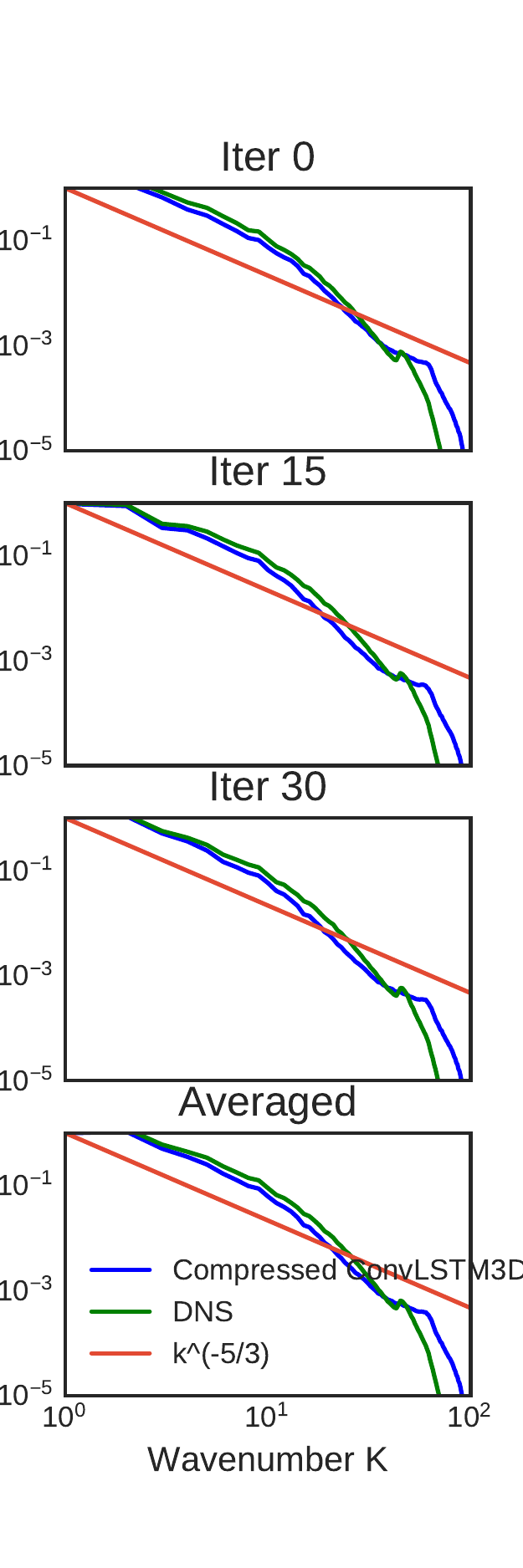}
		\caption{Energy spectra.}
		\end{subfigure}
		\begin{subfigure}{0.18\textwidth}
		\includegraphics[width=\textwidth]{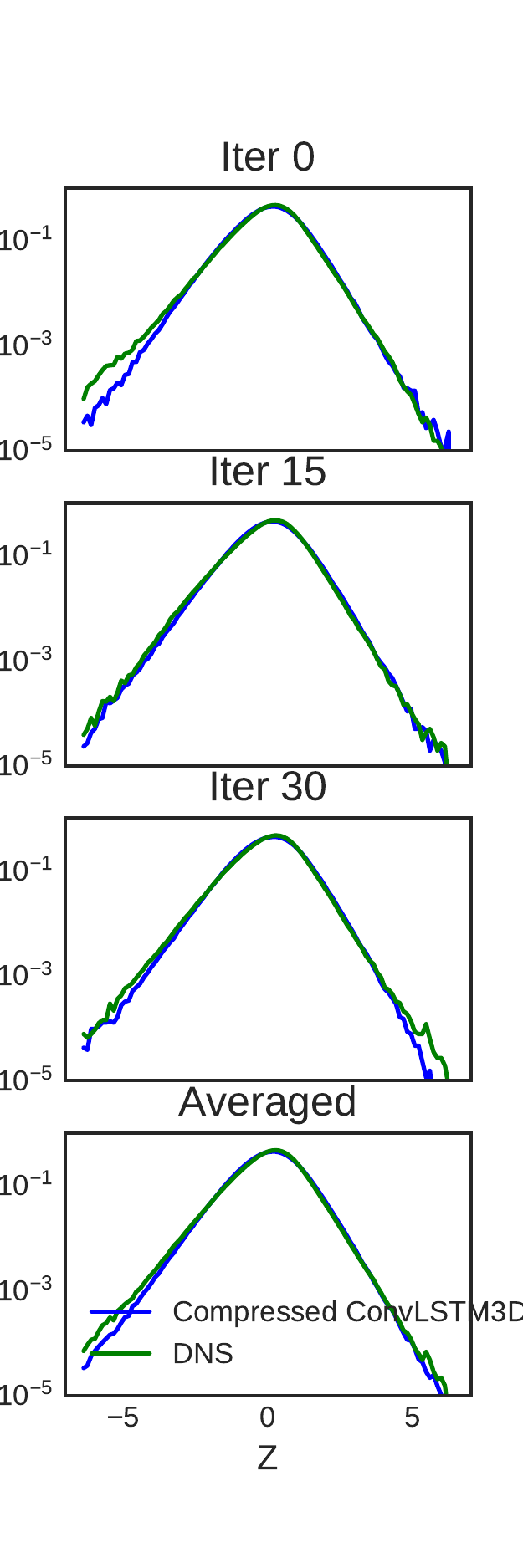}
		\caption{Intermittency.}
		\end{subfigure}
		\begin{subfigure}{0.36\textwidth}
		\includegraphics[width=\textwidth]{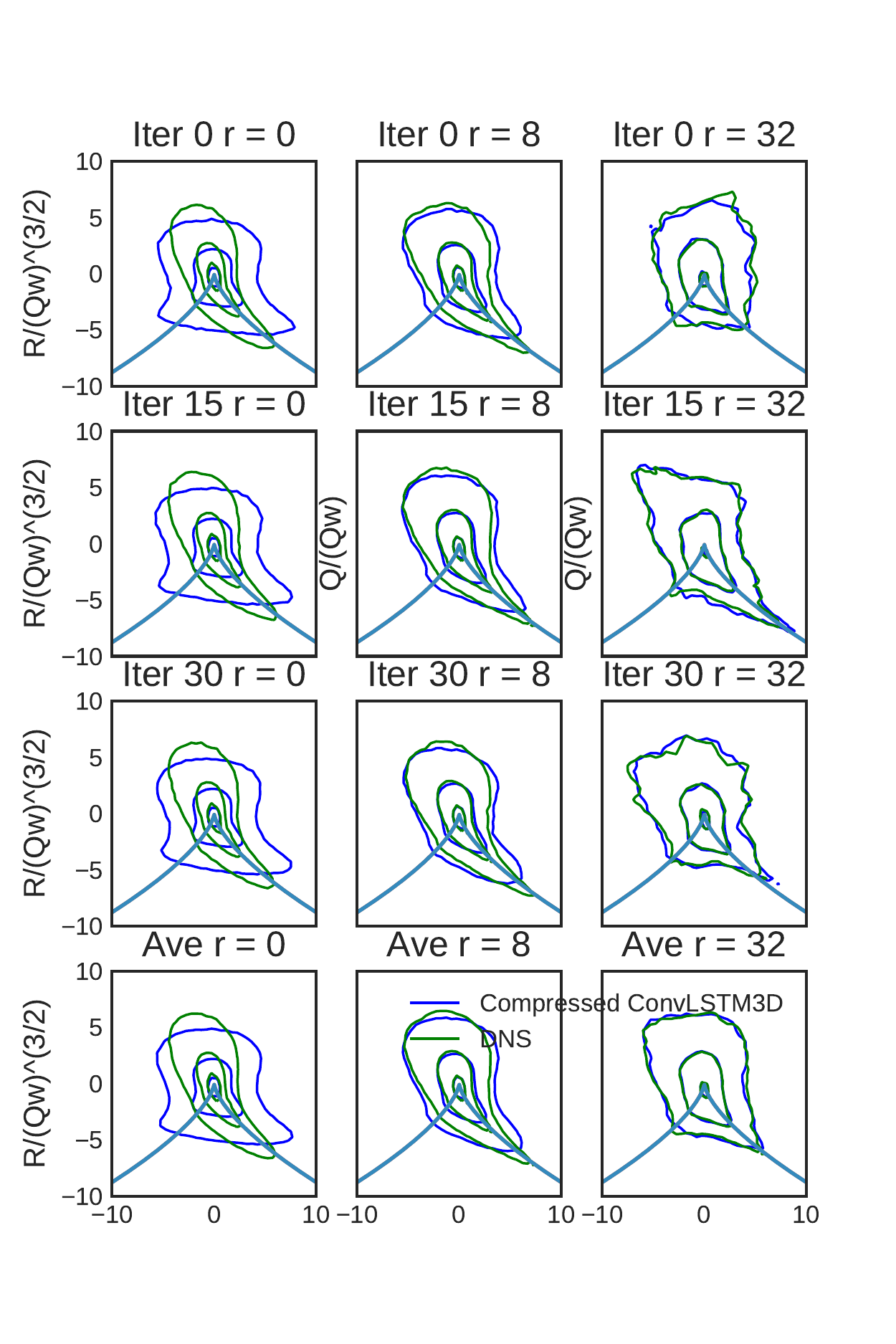}
		\caption{Coarsed-grained Q and R joint PDF's.}
		\end{subfigure}
		\caption{Results of the turbulence diagnostics tests for the dynamic NN LAT-NET (first row) and CC-LSTM (second raw) schemes. Diagnostics is static (applied to 3d snapshots). Notations are equivalent to ones used in Fig.~\ref{fig:GANS}. See also SM B for description of the details.}
		\label{fig:results}
\end{figure}

\emph{LAT-NET.} Dynamic approach to modeling turbulence is to make temporal predictions given an initial state of the flow. Toward this end, the test was a modified version of LAT-NET \citep{Hennigh2017}. LAT-NET works by compressing a snapshot of the flow onto a compact latent space and then mimicking the dynamics of the flow on this latent space. Doing this allows the network to generate the flow simulation for significantly less memory and computation then the flow solver.  A convolutional autoencoder is used to create the mapping to and from the latent space while a separate convolutional network is applied iteratively on the latent space to reproduce the dynamics of the underlying flow. Fig.~\ref{fig:LatNet+CC-LSTM} presents the scheme of operations where the mapping on the latent space is referred to as the latent mapping.  Each application of the latent mapping is meant to ditto a fixed amount of time in the simulation. In our case, each compression mapping corresponds to $0.06$ sec in the simulation thus applying the mapping $10$ times propagates the predicted flow forward in time $0.6$ seconds. The Lat-Net method was first proposed to predict lattice Boltzmann fluid simulations however we have relaxed this constraint so that the network can be used on flow data regardless of the underlying solver. We have also removed the boundary conditions as our tests are on homogeneous flow with no physical boundaries. Training the network is also modified from the original work by splitting the training process into two phases. First, the autoencoder is trained on snapshots of the data and then frozen. Next, the latent mapping is trained on the compressed data in a sequential fashion in Fig.~\ref{fig:LatNet+CC-LSTM}.  Breaking up the training process both speeds up learning and reduces the memory usage allowing larger domains to be trained on. Results of the four diagnostic tests, described above and SM B, applied to the dynamic NN LAT-NET are shown in Fig.~\ref{fig:results}.

\emph{Compressed Convolutional LSTM.} Here we describe another dynamic scheme based on the Convolutional LSTM by Shi~\citep{convlstm}, exploiting the strength of the convolutional networks (which captures spatial features) and Long Short Term Memory i.e. LSTM networks (which capture temporal patterns) together in a combined architecture, and also extending it to the case of 3d images. To reduce dimensionality of the data we also compress and decompress input and output images, respectively, by means of two 3d convolutional autoencoders.  The resulting Compressed Convolutional LSTM (CC-LSTM) architecture is illustrated in Fig.~\ref{fig:LatNet+CC-LSTM}. The results from the CC-LSTM scheme can be seen in Fig.~\ref{fig:results}. Results for energy spectra and for statistics of the velocity gradient show good-to-reasonable consistency between the synthetic and training data.
QR diagnostics show good reproduction by the synthetic data of the training data trends at the larger scales. Respective deviations are significant at the smaller scales -- this is a parasitic effect attributed to insufficient spatial resolution of the spatio-temporal (thus more demanding then pure spatial) schemes. 

\section{Conclusions and Path Forward}
\label{sec:conclusions}

In this manuscript we tested three DL schemes in their ability to reproduce turbulent flows. We introduce four tests of increasing complexity. Our first scheme, implementing C-GAN, is static, i.e. it takes as an input only instantaneous snapshots of the turbulence field, completely ignoring any dynamical aspects of turbulence. The scheme paths our first test, testing distribution of energy over scales, with flying colors. The second test, checking statistics of the velocity gradient which is a small scale object, reveals that GAN underestimates tails of the PDF of the velocity gradient, making generated samples less intermittent then the DNS input.  Third test, of the intermittency (non-Gaussianity) at the intermediate scales (from inertial range of turbulence) via analysis of velocity increments, was passed by the C-GAN scheme well. Finally, the most elaborate fourth test, consisting in checking statistics of the coarse-grained velocity tensor, reveals a difficulty of GAN to reproduce details of the vorticity-strain alignment at the largest (energy containing) scales of turbulence. Based on the analysis of the static NN scheme we naturally came to the question -- if dynamic NNs, trained on temporal sequence of snapshots, is capable of correcting for the static scheme deficiencies in reproducing (a) intermittency of the smallest (close to viscous) scales, and (b) statistics of the vorticity-strain alignment of the largest (close to the energy containing) scales? 

To resolve the question we analyzed the dynamic LAT-NET and CC-LSTM NN schemes trained on temporal sequence of turbulence snapshots. We observed that the dynamic schemes perform on the three-test diagnostics (excluding the anomalous scaling test) at least as well as the static NN scheme. Moreover, we discovered that the small scale intermittency (feature (a) above) is now reproduced well, therefore correcting for the caveat of the static NN.  We attribute this success of the dynamic schemes to the direct cascade nature of turbulence -- turbulent dynamics makes small scale statistics universal, i.e. minimally sensitive to larger scale details of the vorticity-strain alignment. We also note that CC-LSTM is reproducing geometry of the flow better at the larger scales than at the smaller scales,  while LAT-NET show the opposite tendency.  Sub-optimal performance of the two dynamic schemes in the anomalous scaling test is attributed to problems with insufficient spatial resolution. This analysis suggests that dynamic networks bring in additional information which improves reproduction of intermittency at smaller scales and geometry of the flow at large scales.

%\bibliographystyle{unsrt}
%\bibliography{Bib/DeepLearning,Bib/LES,Bib/TurbulenceLearning,Bib/turbulence,Bib/PIML_position,Bib/extra}

\appendix 

\newpage 

\section*{Supplementary Materials (SM) for ``From Deep to Physics-Informed Learning of Turbulence: Diagnostics"}

In the two supplemental Sections we give some additional details on the general Physics Informed Machine Learning approach in Section \ref{sec:PIML},  this manuscript is the first step of, and then in the Section \ref{sec:TurbMetric} we present the primer on physics rationale behind the turbulence metric used to diagnose the three ML schemes in the main part of the manuscript.

\section{Physics Informed Machine Learning: The Methodology} 
\label{sec:PIML}

\begin{figure}[!h]
\centering
 \includegraphics[width=1.35\textwidth,page=7]{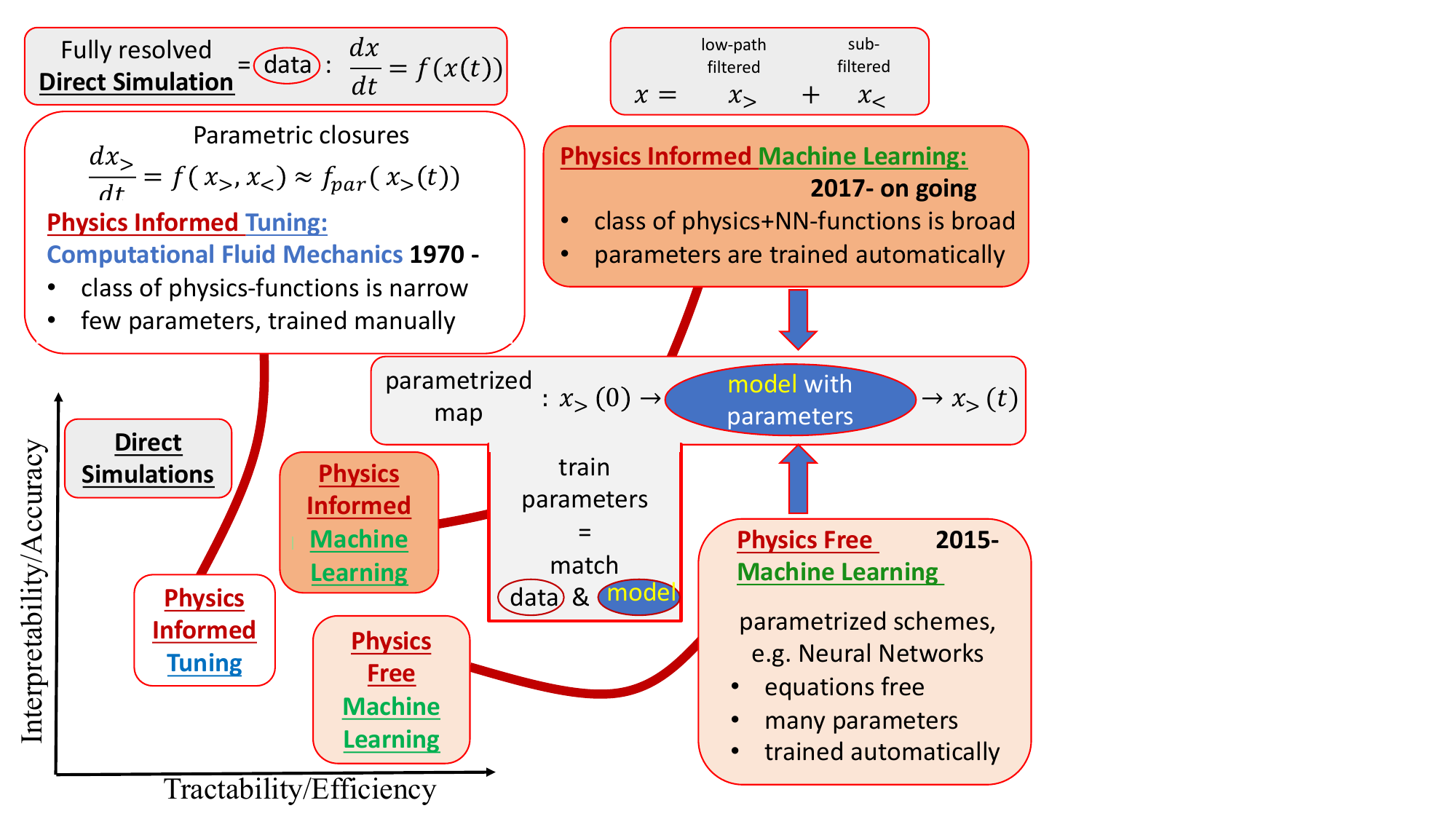}
	\caption{Essence of the PIML framework. \label{fig:PIML}}
\end{figure}

The essence of the Physics Informed Learning (PIML), illustrated in Fig.~(\ref{fig:PIML}), is in capturing physics with the right amount of domain specificity and  interpretability of the Physics Informed Tuning (PIT) non-automatic approach, expressed e.g. in symmetries and constraints,  while also providing prediction power and computational tractability which is on pier with state of the art Machine Learning, such as Deep Learning (DL), which is typically Physics Free, thus PFML.  Development of the PIML approach, which we expect to be transferable to many areas of natural and engineering sciences, is anchored in this manuscript 
%(and more generally in our PIML project at Los Alamos National Laboratory) 
to fluid mechanics. Majority of fluid flows of interest are turbulent, i.e. containing rich spatio-temporal correlations. Goverments, Industry and Science and Engineering communities continue to make significant investments in computational software and hardware to solve turbulence in multiple situations of interest. We conjecture that, when developed, the PIML approach will allow to accelerate turbulent computations by orders of magnitude through separation of scales into macro-scales that are simulated and sub-scales that are efficiently emulated based on PIML learned dynamics. Moreover, the acceleration will be automatic allowing program experts to focus on fewer ad-hoc adjustments of reduced models than is custom today. Also, since a low-dimensional sub-scale state is known the development of closure models (coupling of scales) is greatly simplified. 

In the simplest setting of interest we train parameters of a ML model on the data from Direct Numerical Simulations (DNS) to predict important features of the flow faster than the DNS.  Our main hypothesis is that tremendous acceleration is achievable with PIML. The hypothesis is based on a number of recent reports \citep{King2017,Farimani2017,Miyanawala2017,Hennigh2017,maulik_san_2017,xie2018tempogan,2018WBT,Arvind} that DL schemes developed originally for popular IT industry applications, when applied to DNS data as is, do provide an acceleration. Developing scientific diagnostics for testing and juxtaposing the bare DL approaches and then augmenting DLs based on the results of the diagnostics becomes critical. Complementarily we expect, based on another set of recent publications \citep{Parish2016,Duraisamy2015,Tracey2013,Wang2016,17WWX} that a significant acceleration is also achievable with properly relaxed and parameterized fluid-mechanics phenomenological models, e.g. of Large Eddy Simulations (LES) and Reynolds Averaged Navier Stocks (RANS) types, with much fewer parameters than in the DL models. PIML connects the two approaches to extract robust winning schemes from a number of analyzed synergistic solutions.

Future PIML solutions should be customized to data from DNS of a sequence of turbulence models of increasing complexity, starting from steady homogeneous, isotropic turbulence and extending to non-stationary, non-isotropic turbulent flows involving active (chemical and nuclear) mixing. This strategy will allow us, in particular, to probe and validate transferability of the PIML models trained on simpler cases to more complex situations. 

\section{Turbulence Metric}
\label{sec:TurbMetric}

In this Section we review basic statistical concepts commonly used in the modern literature to analyze results of theoretical, computational and experimental studies of homogeneous isotropic incompressible turbulence in three dimensions. Combination of these concepts are used in the main part of the manuscript as a metric to juxtapose results of the two (static and dynamic) DL methods.  

We assume that a $3d$ snapshot, or its $2d$ slice, or a temporal sequence of snapshots of the velocity field, ${\bm v}=(v^{(i)}({\bm r})|i=1,\cdots,3)$, is investigated. We will focus here on analysis of static correlations within the snapshots. 
%Analyzing data, input or generated, there are three types of averaging possible. First, one may consider averaging over multiple snapshots,  then analyzing dependence on a particular point not assuming an individual snapshot homogeneity,  i.e. invariance under (a three dimensional) shift of the observation points and not assuming individual snapshot isotropy, i.e. invariance under rotation of the reference frame. Second,  we can choose to work with an individual snapshot assuming homogeneity and isotropy. Third,  working with multiple snapshots we combine first and second.

We consider various objects of interest, e.g. correlation functions of second, third and fourth orders
\begin{eqnarray}
&&  \hspace{-1.2cm} C_2^{(i,j)}({\bf r}_1,{\bf r}_2)=\langle v^{(i)}({\bf r}_1) v^{(j)}({\bf r}_2)\rangle,\label{eq:second}\\
&&  \hspace{-1.2cm} C_3^{(i,j,k)}({\bf r}_1,{\bf r}_2,{\bf r}_3)=\langle v^{(i)}({\bf r}_1) v^{(j)}({\bf r}_2)v^{(k)}({\bf r}_3)\rangle,\label{eq:third}\\
&&  \hspace{-1.2cm} C_4^{(i,j,k,l)}({\bf r}_1,\cdots,{\bf r}_4)=\langle v^{(i)}({\bf r}_1) v^{(j)}({\bf r}_2)v^{(k)}({\bf r}_3)v^{(l)}({\bf r}_4)\rangle.\label{eq:fourth}
\end{eqnarray}
%Note that Eq.~(\ref{isotropy}) guarantees that the first moment is zero thus dropped from the list above.
Rich tensorial structure of the correlation functions carry a lot of information. It also suggests a number of useful derived objects, each focusing on a particular feature of the turbulent flows. In particular, we may be to discuss structure functions, defined as tensorial moments of the increments between two points separated by the radius-vector ${\bf r}$:
\begin{eqnarray}
 %S_1^{(i)}({\bm r}) &=&\langle (v^{(i)}({\bf r})-v^{(i)}({\bf 0}))\rangle, \label{S1}\\
 && \hspace{-1.5cm} S_2^{(i,j)}({\bm r})=\langle (v^{(i)}({\bf r})-v^{(i)}({\bf 0}))(v^{(j)}({\bf r})-v^{(j)}({\bf 0}))\rangle, \label{eq:S2}\\
&& \hspace{-1.5cm} S_3^{(i,j,k)}({\bm r})=\langle (v^{(i)}({\bf r})-v^{(i)}({\bf 0}))(v^{(j)}({\bf r})-v^{(j)}({\bf 0}))\nonumber\\
&& \times  (v^{(k)}({\bf r})-v^{(k)}({\bf 0}))\rangle, \label{eq:S3}\\
&& \hspace{-1.5cm} S_4^{(i,j,k,l)}({\bm r})=\langle (v^{(i)}({\bf r})-v^{(i)}({\bf 0}))(v^{(j)}({\bf r})-v^{(j)}({\bf 0}))\nonumber\\
&& \times (v^{(k)}({\bf r})-v^{(k)}({\bf 0}))(v^{(l)}({\bf r})-v^{(l)}({\bf 0}))\rangle. \label{eq:S4}
\end{eqnarray}
Other objects of interest,  derived from the correlation function by spatial differentiation and then merging the points are moments of the velocity gradient tensor, $m^{(i,j)}=\nabla^{(i)} v^{(j)}$
\begin{eqnarray}
&& \hspace{-1cm} D_2^{(i_1,i_2;j_1,j_2)}=\langle m^{(i_1,j_1)}m^{(i_2,j_2)}\rangle, \label{eq:grad2}\\
&& \hspace{-1cm} D_3^{(i_1,i_2,i_3;j_1,j_2,j_3)}=\langle m^{(i_1,j_1)}m^{(i_2,j_2)}m^{(i_3,j_3)}\rangle, \label{eq:grad3}\\
&& \hspace{-1cm} D_4^{(i_1,i_2,i_3,i_4;j_1,j_2,j_3,j_4)}=\langle m^{(i_1,j_1)}m^{(i_2,j_2)}m^{(i_3,j_3)} m^{(i_3,j_3)}\rangle. \label{eq:grad4}
\end{eqnarray}
We may also be interested to study mixed objects, e.g. the so-called energy flux
\begin{eqnarray}
F({\bf r})=\langle v^{(j)}({\bf 0}) v^{(i)}({\bf r}) m^{(i,j)}({\bf r})\rangle.
\label{eq:flux}
\end{eqnarray}

The remainder of this Section is organized as follows. We describe important turbulence concepts mentioned in the main part of the paper one by one,  starting from simpler ones and advancing towards more complex concepts. 

\paragraph{$4/5$ Kolmogorov law and the Energy Spectra}

Main statement of the Kolmogorov theory of turbulence (in fact it is the only formally proved statement of the theory) is that asymptotically in the inertial range, i.e. at $L\gg r\gg\eta$, where $L$ is the largest, so-called energy-containing scale of turbulence and $\eta$ is the smallest scale of turbulence, so-called Kolmogorov (viscous) scale, $F(r)$ does not depend on $r$.  Moreover, the so-called $4/5$-law states for the third-order moment of the longitudinal velocity increment
\begin{eqnarray}
&& L\gg r\gg\eta:\quad S_3^{(i,j,k)}\frac{r^i r^j r^k}{r^3}=-\frac{4}{5}\varepsilon r,
\label{eq:4/5}
\end{eqnarray}
where $\varepsilon=\nu D_2^{(i,j;i,j)}/2$ is the kinetic energy dissipation also equal to the energy flux.

Self-similarity hypothesis extended from the third moment to the second moment results in the expectation that within the inertial range, $L\gg r\eta$, the second moment of velocity increment scales as, $S_2(r)\sim v_L (r/L)^{2/3}$. This feature is typically tested by plotting the energy spectra of turbulence (expressed via $S_2(r)$) in the wave vector domain, e.g. as shown in Figs.~(1a,3a) of the main text. 

\paragraph{Normal and Anomalous scaling of velocity increments}

One expects moments of the velocity increment to show the following scaling behavior inside the inertial range of turbulence
\begin{eqnarray}
&& L\gg r\gg\eta:\quad S_n(r)\sim (v_L)^n\left(\frac{r}{L}\right)^{n/3-\Delta_n}
\label{eq:S_n_scaling}
\end{eqnarray}
where $L$ is the energy containing (largest) scale of turbulence, $\eta\sim (\nu/v_L)^{3/4}L^{1/4}$ is the Kolmogorov (viscous) scale, $v_L$ is the typical velocity fluctuations at the energy containing scale, $\nu$ is the viscosity coefficient and $\Delta_n$ is the so-called anomalous scaling. $\Delta_3=0$, and $\Delta_n$ is a growing function of $n$. In 1941 A. Kolmogorov \citep{41Kol_a,41Kol_b,41Kol_c} hypothesized self-similarity of turbulence, i.e. that the anomalous scaling is absent, $\forall n=3,\cdots:\quad \Delta_n=0$.  Criticized by  L. Landau (see e.g. \citep{11Fal} for related history notes), Kolmogorov reconsidered self-similarity in 1962 \citep{62Kol},  suggesting instead the so-called refined (log-normal) similarity hypothesis, $\Delta_n=\mu n (n-3)/16$,  which was then confirmed to be rather accurate in experiments and simulations, see e.g. \citep{95Fri}. 

The intermittency test was implemented for both static and dynamic NN schemes discussed in the main part of the manuscript. Both static and dynamic NN schemes passed the test sucessfully.  We do not show results in the main part of the manuscript because of the space limitations. 

\paragraph{Intermittency of Velocity Gradient}

Consistently with Eq.~(\ref{eq:S_n_scaling}), estimation of the moments of the velocity gradient results in
\begin{eqnarray}
&& D_n\sim \frac{S_n(\eta)}{\eta^n},
\label{eq:D_n_scaling}
\end{eqnarray}
where dependence of the $r/L$- and $v_L$- independent pre-factors in both Eq.~(\ref{eq:S_n_scaling}) and Eq.~(\ref{eq:D_n_scaling}) on $n$ is ignored. Intermittency (extreme non-Gaussianity) of turbulence is expressed the strongest in Eq.~(\ref{eq:D_n_scaling}).

\paragraph{Statistics of coarse-grained velocity gradients: $Q-R$ plane.}

Isolines of probability in the $Q-R$ plane, expressing intimate feature of the turbulent flow geometry, has a nontrivial shape documented in the literature. See \citep{99CPS,11Men} and references therein. Different parts of the $Q-R$ plane are associated with different structures of the flow. Thus lower right corner (negative $Q$ and $R$), which has higher probability than other quadrants, corresponds to a pancake type of structure (two expending directions, one contracting) with the direction of rotation (vorticity) aligned with the second eigenvector of the stress. This tea-drop shape of the probability isoline becomes more prominent with decrease of the coarse-graining scale.

\end{document}